\documentclass[useAMS,usenatbib]{mn2e}
\usepackage{epsfig, journals}
\newcommand{\citenp}[1]{\citeauthor{#1} \citeyear{#1}} 

\newcommand{\scri}{\scriptsize} 

\newcommand{\vc}[1]{\mbox{\bf #1}}

\newcommand{\tfrac}[2]{{\textstyle\frac{#1}{#2}}}

\newcommand{\sgn}{\mbox{sgn}\,}
\newcommand{\etaup}{\eta_{\uparrow}}
\newcommand{\etadown}{\eta_{\downarrow}}

\title[Interface dynamo with meridional flow]{The effect of a meridional flow 
on Parker's interface dynamo}
\author[K. Petrovay and A. Kerekes]{K. Petrovay$^{1}$\thanks{E-mail:
K.Petrovay@astro.elte.hu}, and A. Kerekes$^{2}$\\
$^{1}$E\"otv\"os University, Department of Astronomy, Budapest, Pf. 32, 
    H-1518 Hungary\\
$^{2}$SPARC, Dept. of Applied Mathematics, Univ. of Sheffield, 
Hicks Building, Hounsfield Road, Sheffield S3 7RH, UK}

\begin{document}

\date{MNRAS, in press}

\pagerange{\pageref{firstpage}--\pageref{lastpage}} \pubyear{2004}

\maketitle

\label{firstpage}

\begin{abstract}
Parker's interface dynamo is generalized to the case when a homogeneous flow is
present in the high-diffusivity (upper) layer in the lateral direction (i.e.\
perpendicular to the shear flow in the lower layer). This is probably a
realistic first representation of the situation near the bottom of the solar
convective zone, as the strongly subadiabatic stratification of the tachocline
(lower layer in the interface dynamo) imposes a strong upper limit on the
speed of any meridional flow there.

Analytic solutions to the eigenvalue problem are presented for the cases of
vanishing diffusivity contrast and infinite diffusivity contrast, respectively.
Unlike the trivial case of a homogeneous system, the ability of the meridional
flow to reverse the propagation of the dynamo wave is strongly reduced in the
interface dynamo. In particular, in the limit of high diffusivity contrast
relevant to the solar case it is found that a meridional flow of realistic
amplitude cannot reverse the direction of propagation of the dynamo wave. The
implications of this result for the solar dynamo problem are discussed.
\end{abstract}


\section{Introduction}

While the operation of the solar dynamo is still far from understood
(\citenp{Weiss:CUPrev}, \citenp{Petrovay:SOLSPA}), it is now generally believed
that the strong toroidal magnetic field responsible for solar activity is
generated and stored in the tachocline layer. This transitional layer between
the differentially rotating convective zone (CZ) and the rigidly rotating solar
interior is characterized by a strong radial shear, and is thus an ideal
candidate for the production of the toroidal field. 

The site and physical nature of the toroidal$\rightarrow$poloidal flux
conversion ($\alpha$-effect), needed to close the cycle, is much less clear.
One popular possibility is the interface dynamo
(\citenp{Parker:interface}---hereafter P93,\citenp{Tobias:nonlin.interface}, 
\citenp{Charbonneau+McGregor:IFdynamo}, \citenp{Markiel+Thomas},
\citenp{Mason+:competition}) where  $\alpha$ is concentrated in the deepest
part of the CZ, immediately above the layer of strong shear.

Helioseismic inversions (\citenp{Kosovichev:tachocline};
\citenp{Basu+Antia:tachovar}) indicate that at low latitudes the solar
tachocline lies immediately below the adiabatically stratified CZ. This implies
that the turbulent diffusivity in the tachocline is significantly suppressed in
comparison to the layers immediately above it, resembling the situation
envisaged in interface dynamo models.  An attractive feature of these models is
that the sharp diffusivity contrast between the two layers gives rise to quite
strong toroidal magnetic fields, in agreement with the requirements of flux
emergence calculations. If the sign of $\alpha$ is negative on the northern
hemisphere, as expected near the bottom of the CZ, then the Parker--Yoshimura
sign rule predicts an equatorward propagating dynamo wave at low latitudes and
a poleward propagating wave at high latitudes, with the two belts departing at
the corotation latitude of $\sim35^{\circ}$. It is tempting to identify these
waves with the two well known branches of the extended butterfly diagram (e.g.\
\citenp{Makarov+Sivaraman}; cf. also \citenp{Petrovay+Szakaly:2d.pol}).

The presence of a meridional circulation, however, significantly complicates
the picture. As a poleward meridional flow of amplitude $\sim 10$--$20\,$m/s is
clearly detected in the shallower layers of the CZ
(\citenp{Hathaway:merid.flow}), continuity seems to require a counterflow in
the deep CZ (though a two-cell pattern is at present also not
excluded). As this velocity amplitude is comparable to the speed of migration
of the activity belts, the question arises whether the meridional flow can
invalidate the Parker--Yoshimura sign rule, reversing the propagation of a
dynamo wave. Indeed, in flux transport models of the solar dynamo
(\citenp{Wang+:1.5D}; \citenp{Choudhuri+:mixed.transp}), a deep equatorward
meridional flow is responsible for the migration of sunspot-forming latitudes
during the solar cycle.

As the effect of a homogeneous flow can obviously be described by a Galilean
transformation of the solution with no flow, it is indeed to be expected that
an equatorward meridional flow pervading the whole dynamo region near the
bottom of the CZ (as assumed in flux transport models) will reverse the
poleward propagation of the dynamo wave if its speed is higher than the phase
velocity. (In these models, $\alpha$ is concentrated near the surface and is
positive in the northern hemisphere, leading to poleward propagation at low
latitudes ---hence the need for a reversal.) The actual situation is, however,
more complex, as a meridional flow of significant amplitude cannot be expected
to penetrate below the adiabatically stratified CZ. The strong subadiabatic
stratification of the tachocline represents a serious obstacle in the way of
meridional circulation. The timescale of any meridional flow here cannot be
shorter than the (turbulent) heat diffusion  timescale, allowing downmoving
fluid elements to get rid of their strong  buoyancy. Heat diffusivity in the
subadiabatic layer is clearly much lower than above, so the same must be true
for the amplitude of the poloidal flow. A more realistic representation of the
meridional flow pattern in the dynamo layer, then, is an interface-type model
with no meridional flow in the strongly sheared, low-diffusivity lower layer,
and a homogeneous meridional flow in the highly diffusive upper layer. 

This paper addresses the problem of how a meridional flow influences the
properties of the dynamo wave by considering the arguably simplest nontrivial
case: the effect of a homogeneous meridional flow in the top layer of Parker's
 interface dynamo. Section 2 describes the analytic
model, Section 3 presents some solutions for important special cases. Finally,
in Section 4 the implications of our findings are discussed for the problems
outlined above.

\section{Model and Dispersion Relation}

Our model is a generalization of Parker's Cartesian interface dynamo
(P93). The model describes dynamo activity occurring in
two adjacent layers, say $z<0$ and $z>0$. This Cartesian model can be
considered a local approximation of the true spherical system, with the $x$ and
$y$ coordinates  corresponding to colatitude and longitude, respectively.
The magnetic field is written in the form 
\[
\vc B=B\vc e_y + \nabla\times(A\vc e_y),
\]
while the zonal flow is written in terms of a shear rate $\Omega$:
\[ u_y=\Omega z  \]
The parameters are homogeneously distributed in each layer, and change
discontinuously across the interface. In the upper part,  we assume $u_x=u_0$,
$\Omega=0$, $\alpha=\alpha_0$, $\eta=\etaup$,  while in the lower part, we have
$\Omega=\Omega_0$, $\alpha=u_x=0$, $\eta=\etadown$. (Here and in what follows,
the indices $\uparrow$ and $\downarrow$ refer to values in the $z>0$ and $z<0$
sublayers, respectively.)

The $\alpha\Omega$-dynamo equations in the two layers read
\begin{equation}
  \partial_t A + u_0\partial_x A =\eta \nabla^2 A + \alpha B,
  \label{eq:poldynamo}
\end{equation}
\begin{equation}
  \partial_t B + u_0\partial_x B =\eta \nabla^2 B + \Omega \partial_x A
  \label{eq:tordynamo}
\end{equation}
with the appropriate boundary conditions ensuring conservation of magnetic flux
applied at the interface ($z=0$).

Following P93, the solution of equations
(\ref{eq:poldynamo}) and (\ref{eq:tordynamo}) is now sought in the form
\begin{equation}
\begin{array}{ll}
B_\uparrow= C\,\exp(\sigma t-Sz)\,\exp[i(\omega t+kx-Qz)]  \\
&\\
B_\downarrow= (L+Mz)\,\exp(\sigma t+sz)\,\exp[i(\omega t+kx+qz)]  \\
&\\
A_\uparrow= (D+Ez)\,\exp(\sigma t-Sz)\,\exp[i(\omega t+kx-Qz)]  \\
&\\
A_\downarrow= J\,\exp(\sigma t+sz)\,\exp[i(\omega t+kx+qz)]
\end{array}
\label{eq:ansatz}
\end{equation}
where all parameters are real. At the interface $z=0$, we require
\begin{equation}
\begin{array}{cc}
A_\downarrow=A_\uparrow, \quad  \partial_z A_\downarrow=\partial_z A_\uparrow, 
  \quad B_\downarrow=B_\uparrow,  \quad 
  \etadown \partial_z B_\downarrow  =\etaup \partial_z B_\uparrow. 
\end{array}
%
%\begin{array}{cc}
%A_\downarrow=A_\uparrow, & B_\downarrow=B_\uparrow,                       \\
%&\\
%\displaystyle \partial_z A_\downarrow=\partial_z A_\uparrow, & 
%\displaystyle\quad \etadown \partial_z B_\downarrow 
%   =\etaup \partial_z B_\uparrow. 
%\end{array}
\label{eq:bcond}
\end{equation}

Nondimensional parameters are now introduced as
\begin{equation}
\begin{array}{ll}
\mbox{diffusivity contrast:} & \mu^2=\etadown/\etaup, \\
\mbox{dynamo number:} & N=\alpha_0\Omega_0/\etaup^3 k^3, \\
\mbox{Reynolds number:} & R=u_0/\etaup k, \\
\mbox{dimensionless frequency:} & \nu=\omega/\etaup k^2,  \\
\mbox{dimensionless growth rate: } \qquad & \beta-1=\sigma/\etaup k^2.
\end{array}
\label{eq:params}
\end{equation}
Together with equation (\ref{eq:ansatz}), this implies that the  nondimensional
phase speed of a mode is $-\nu$.

Substituting the {\it ansatz} (\ref{eq:ansatz}) into equations (4) and
(5), taking into account the conditions (\ref{eq:bcond}), and following
the procedure of P93, we find the complex dispersion 
relation
\begin{eqnarray}
&&
Z^8(\mu^2-1)^2+2Z^6(\mu^2-1)^2(\mu^2-1-iR)\nonumber\\&&
+Z^4[(1-3\mu^2+\mu^4)(\mu^2-1-iR)^2-i(1+\mu^2)\mu^2N/2] \nonumber\\&&
-Z^2\mu^2[(\mu^2-1-iR)^3+i(1+\mu^2)(\mu^2-1-iR)N]
\nonumber\\&&
-\mu^4N^2/16=0,
\label{eq:disp}
\end{eqnarray}
with
%\begin{equation}
$Z^2=\beta+i(\nu+R)$. 
%\end{equation}
By separating equation (\ref{eq:disp}) into real and imaginary parts, one
arrives at the dispersion relations to be solved for the real quantities
$\beta$ and  $\nu$.  While these relations can be given in a closed form, they
are rather lengthy and awkward to handle in the general case. However, the
dispersion  relations simplify significantly for some important special cases.
In what follows, we will consider each of these cases in turn.

\section{Solutions for Special Cases}

\subsection{Results for the Case $\mu=1$}

The real and imaginary parts of the dispersion relation (\ref{eq:disp}) now
simplify to
\begin{equation}  
\beta ^2 R^2 + \beta N(R + 2\nu ) - \nu R^2(R + \nu ) - N^2/16 =0,
\label{eq:mu1disp1}
\end{equation}
\begin{equation}  
\beta^2 N + \beta R^2(R+2\nu ) - \nu N(R+\nu )= 0.
\label{eq:mu1disp2}
\end{equation}
Multiplying equation (\ref{eq:mu1disp1}) by $N$ and equation (\ref{eq:mu1disp2})
by $R^2$, and taking the difference we have
\begin{equation}  
\beta (R+2\nu )(R^4+N^2) - N^3/16= 0.
\end{equation}
From this, for the frequency $\nu$ we have the explicit result
\begin{equation}
\nu =\frac{N^3}{32\beta (R^4+N^2)} - \frac{R}{2},
\label{eq:mu1nu}
\end{equation}
while for the growth rate $\beta-1$ we a quadratic equation in $\beta^2$:
\begin{eqnarray}
&&N^6 -64 N^4(16\beta ^4 +3R^2\beta ^2) 
-64N^2(32R^4\beta ^4 + 7R^6\beta ^2) 
\nonumber\\&&
- 256 (4R^8\beta ^4 + R^{10}\beta ^2) = 0. 
\label{eq:mu1beta}
\end{eqnarray}

\begin{figure}
\epsfig{figure=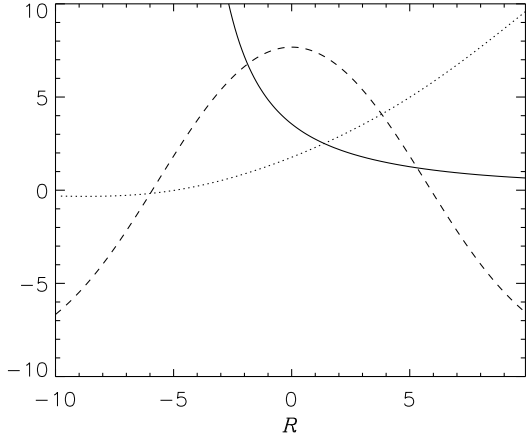,width=84mm}
\caption{Dynamo wave period $P=2\pi/|\nu|$ (solid), phase speed $-\nu$
(dotted), and growth rate $\beta-1$ (dashed), as functions of the flow speed
$R$ in the analytic model with uniform diffusivity $\mu=1$, with $N=-100$. }
\label{fig:mu1}
\end{figure}

\begin{figure}
\epsfig{figure=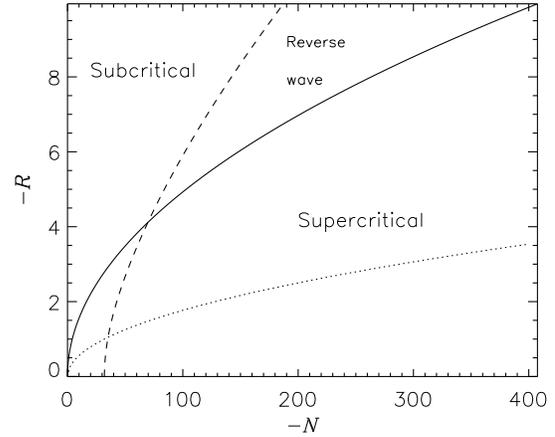,width=84mm}
\caption{Minimum antiparallel flow speed necessary to reverse the dynamo
wave (solid) in the case of uniform diffusivity in the analytical model. Dashed
line indicates the critical mode separating the sub- and supercritical regimes.
The dotted line shows the negative of the phase speed $-\nu$ for $R=0$. In the
case of a uniform flow the solid and dotted lines would coincide.}
\label{fig:turnaround}
\end{figure}

In Figure~\ref{fig:mu1}, the wave period $P=2\pi/|\nu|$, the phase speed $-\nu$,
and the growth rate $\beta-1$, computed from relations (\ref{eq:mu1nu}) and 
(\ref{eq:mu1beta}), are plotted for $N=-100$.  The critical dynamo number for
$R=0$ is $N_{\mbox{\scri cr}}=\pm 32$, so this corresponds to a moderately
supercritical case.  It is apparent that for high negative values of $R$ the
phase speed becomes slightly negative, indicating that the dynamo wave is 
reversed.
Nevertheless, the minimal flow speed needed for this, $R\simeq -5$, is
significantly higher in modulus than the phase speed for $R=0$ ($-\nu\sim 1$),
in contrast to the case of a uniform flow. This confirms that with the
nonuniform flow considered here the properties of the dynamo differ significantly
from the case of a uniform flow that simply advects the dynamo wave. In
particular, the Parker-Yoshimura sign rule
\begin{equation}
\sgn\nu=\sgn N , 
\label{eq:signrule}
\end{equation}
that determines the sense of dynamo wave propagation when $R=0$, is not as
easy  to violate as in the case of a uniform flow. This is further borne out in
Figure~\ref{fig:turnaround}, where the wedge-shaped region marked ``reverse
wave'' denotes that portion of the $N$-$R$ plane in which the sign rule is
violated.  For a uniform flow, this region should clearly extend down to the
dotted line indicating the negative of the phase speed in the $R=0$ case.

It is apparent that the growth rate is maximal for $R=0$, and otherwise
independent  of the direction of the flow (as only even powers of $R$ appear in
equation  [\ref{eq:mu1beta}]).  This symmetric behaviour of the $\beta(R)$
curve is, however, specific to the case $\mu=1$. Indeed, expanding the
dispersion relation in terms of $m=\mu^2-1$ and  $R$ around the case $m=R=0$,
after considerable algebra one finds
\begin{displaymath}
\frac{\partial\beta}{\partial R}{\bigg\vert}_{R=0}= 
\frac{-3 mN}{1024\,\beta_0^3}
\end{displaymath}
$\beta_0-1$ being the growth rate for $m=R=0$. From this it is clear that for
$N<0$, in models with  $\mu^2<1$, the growth rate is increased by a finite
negative  $R$, while with $\mu^2>1$ the reverse is true. 

\subsection{Results for the Case $\mu\ll 1$}

We now turn to the case more relevant to the Sun, when the shear layer is
characterized by a lower value of the diffusivity than the layer above. Again
following P93, we consider the limit $\mu\rightarrow 0$ while $\tilde
N=\mu^2 N$ remains finite, and restrict our attention to unstable modes with
$\mu^2\ll\beta-1$. The real and imaginary parts of the dispersion relation
(\ref{eq:disp}) can then be written as
\begin{eqnarray}
&&
(\nu+R)^4-2R(\nu+R)^3-(\nu+R)^2(6\beta^2-6\beta+1-R^2) 
\nonumber\\&&
+(\nu+R)[2\beta R(3\beta-2)+\tilde N(\beta-\tfrac 12)] +\beta^2(\beta-1)^2
\nonumber\\&&
-\beta R(\beta R+\tfrac 12\tilde N) -\tfrac 1{16}{\tilde N}^2 =0,
\label{eq:mu0disp1}
\end{eqnarray}
\begin{equation}
[(\nu+R)^2-\beta(\beta-1)-(\nu+R)R][\tilde N-8(\nu+R)(\beta-\tfrac 12)+4\beta R]
=0.
\label{eq:mu0disp2}
\end{equation}
The latter relation implies that either
\begin{equation}
\nu=\tfrac 12\{-R\pm[R^2+4(\beta-1)\beta]^{1/2}\},
\label{eq:mu0nu}
\end{equation}
or
\begin{equation}
\tilde N=8(\nu+R)(\beta-\tfrac 12)-4\beta R.
\label{eq:mu0nu2}
\end{equation}
Substituting either of the last two relations into (\ref{eq:mu0disp1}) yields
the other, so by combining them we obtain a relation that gives the growth
rate implicitly
\begin{equation}
\tilde N=4\{R\pm[R^2+4(\beta-1)\beta]^{1/2}\}(\beta-\tfrac 12) -4\beta R.
\label{eq:mu0beta}
\end{equation}
Upon solving this for $\beta$, $\nu$ can be calculated from equation
(\ref{eq:mu0nu2}). It is clear that for $\tilde N>0$ it is the $+$ sign, while
for $\tilde N<0$ it is the $-$ sign that corresponds to growing modes in
the case $R=0$. The other sign will not lead to growing modes for any value of
$R$.

\begin{figure}
\epsfig{figure=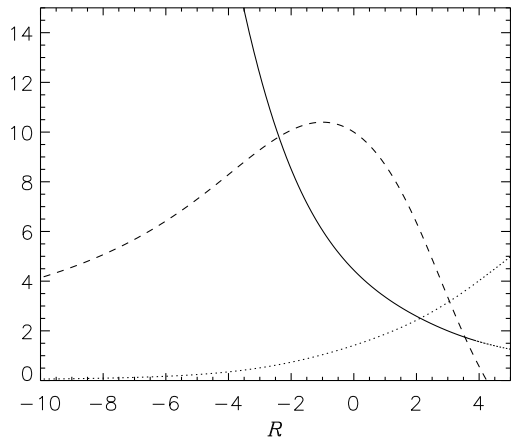,width=84mm}
\caption{Dynamo wave period $P=2\pi/|\nu|$ (solid), phase speed $-\nu$
(dotted), and growth rate $\beta-1$ (dashed), as functions of the flow speed
$R$ in the analytic model in the limit of very high diffusivity contrast 
$\mu^2\ll\beta-1$, with $N=-12\sqrt 2$. }
\label{fig:mu0}
\end{figure}

The dependence of the wave period $P=2\pi/|\nu|$, the phase speed $-\nu$, and
the growth rate $\beta-1$ on $R$ is plotted in Figure~\ref{fig:mu0} for the
growing mode ($-$ sign), in a moderately supercritical case with a negative
dynamo number.  

Inspection of Figure~\ref{fig:mu0} and, more importantly, the analysis of
equations (\ref{eq:mu0nu})-(\ref{eq:mu0beta}) also indicates that $-\nu>0$ for
all values of $R$ for growing modes; that is, the sign rule (\ref{eq:signrule})
seems to hold unconditionally in the limit considered here. Indeed, while a
parallel flow of relatively moderate speed $R_{\mbox{\scri max}}$ ($\sim 4$ in
the case considered) can quench the dynamo completely, an antiparallel flow
($R<0$) can
apparently neither quench the dynamo wave, nor turn it around, for any finite
speed.  However, a caveat is in order when applying these findings to the case 
of a small but finite $\mu$. As the results were derived assuming
$\mu^2\ll\beta-1$, their applicability is limited to a limited range in  $R$,
the limits of which are determined by the condition $\beta(R)-1\gg\mu^2$. 
%In fact, it is possible to show that for a finite, though rather high, 
%velocity, an antiparallel flow can ultimately reverse, and then quench the 
%dynamo wave, in agreement with the numerical results of Figure~7 (curves $c$). 
However, from equation (\ref{eq:mu0beta}) in the case of an antiparallel flow 
one finds the condition $R\gg R_{\mbox{\scriptsize cr}}=\tilde N/4\mu^2$ (for
$\tilde N<0$), the order of  magnitude of $R_{\mbox {\scriptsize cr}}$ being at
least 100, much higher than any realistic estimate of $R$ in the solar case.
Thus we can safely conclude that {\it in the case of high diffusivity contrast
and meridional flow limited to the upper regime (CZ), a meridional
flow cannot reverse the direction of propagation of the dynamo wave,} unless
its amplitude is unrealistically high.

%\begin{figure}
%\epsfig{figure=4abra.eps,width=84mm}
%\caption{Poloidal field lines at an arbitrary
%instant for different values of $R$, in the case $\tilde N=50$, 
%$\mu^2=0.01$.}
%\label{fig:poloidal}
%\end{figure}

%Finally, in Figure~\ref{fig:poloidal} we present the field lines of the
%poloidal field at an arbitrary instant, for different values of the meridional
%flow speed, this time for a case with positive dynamo number (i.e. negative
%phase speed). It is apparent that an increasing positive (rightwards directed)
%meridional ``wind'' in the upper regime ``blows'' the peripheric parts of the
%leftwards  migrating field, generated near the interface, ever more effectively
%back.

\section{CONCLUSION}

We have studied by analytical methods how a meridional flow
limited to the upper (i.e.~unsheared) layer affects kinematic interface
dynamos. We found that the growth rate and period of the dynamo wave have
a nontrivial dependence on the flow amplitude. In the case of strongly reduced
diffusivity in the lower layer, relevant for the Sun, a flow parallel to the
direction of propagation of the dynamo wave reduces the growth rate until, at a
finite critical flow speed, it completely suppresses dynamo action. An
antiparallel flow, in contrast, first increases the growth rate; then, after
reaching a maximum, the growth rate tends to zero as $u_0\rightarrow-\infty$.
Contrary to intuition, an antiparallel flow can neither suppress nor revert the
direction of propagation of the dynamo wave.

This conclusion is apparently at odds with flux transport models
(\citenp{Choudhuri+:mixed.transp}, \citenp{Dikpati+Charbonneau}) where an
equatorward flow near the bottom of the CZ can turn the direction of
propagation of the dynamo wave towards the equator at low latitudes. There is,
however, no real contradiction here, as in those models the flow is assumed to
penetrate quite deep into the layer of strong shear. In terms of the simple
Cartesian model studied here, a closer counterpart of the flux transport models
would be a case with $u_0=$const. throughout the region: then an antiparallel
flow exceeding the phase speed can trivially turn around a dynamo wave. In
Section~1 we argued that it is more realistic to assume that the meridional
flow is limited to the adiabatic upper layer. In fact, the more recent flux
transport model of \citet{Dikpati+:polar.fields} admits that the circulation
must be limited to the adiabatic layer. Therefore, in order to make the model
work, they need to assume that a significant fraction of the tachocline
overlaps the adiabatic SCZ. This may indeed hold at high latitudes, but at low
latitudes, where a meridional transport of toroidal fields is most needed, it
does not seem to be supported by helioseismic data 
(\citenp{Basu+Antia:tachovar}).

These arguments indicate that our simplified models may have more relevance to
the solar dynamo than many complex nonlinear spherical models where the assumed
geometrical distribution of the flows and transport coefficient does not
reflect the situation in the solar interior. Nevertheless, owing to the
simplified  2D Cartesian geometry and the kinematical nature of our models our
conclusions should be taken with proper reservation. In this context it may be
mentioned that that details of the physical structure, such as the precise form
of the rotation law, were shown to have a profound effect on the behaviour of
dynamo models (\citenp{Moss+:towards}, \citenp{Phillips+:dyn.structure}). 
Extensions of this work to spherical geometry and to the 3D nonlinear domain
are clearly needed to tell whether the present results remain valid in more
realistic situations.

\section*{Acknowledgments}
This work was supported in part by the OTKA under
grants no.\ T034998 and T043741 and by the Royal Society grant no.~15599.

%\bibliographystyle{mn2e}
%\bibliography{kris}

\label{lastpage}

\end{document}